# Three dimensional photonic Dirac points in metamaterials

Qinghua Guo[1,2], Biao Yang[2], Lingbo Xia[2,3], Wenlong Gao[2], Hongchao Liu[2], Jing Chen[4], Yuanjiang Xiang[1*], Shuang Zhang[2†]

[1]*Key Laboratory of Optoelectronic Devices and Systems of Ministry of Education and Guangdong Province, College of Optoelectronic Engineering, Shenzhen University, Shenzhen 518060, China*

[2]*School of Physics and Astronomy, University of Birmingham, Birmingham B15 2TT, United Kingdom*

[3]*Center for Terahertz Waves and College of Precision Instrument and Optoelectronics Engineering, Tianjin University, Tianjin 300072, China*

[4]*MOE Key Laboratory of Weak-Light Nonlinear Photonics, School of Physics, Nankai University, Tianjin 300071, China*

**Abstract:** Topological semimetals, representing a new topological phase that lacks a full bandgap in bulk states and exhibiting nontrivial topological orders, recently have been extended to photonic systems, predominantly in photonic crystals and to a lesser extent, metamaterials. Photonic crystal realizations of Dirac degeneracies are protected by various space symmetries, where Bloch modes span the spin and orbital subspaces. Here, we theoretically show that Dirac points can also be realized in effective media through the intrinsic degrees of freedom in electromagnetism under electromagnetic duality. A pair of spin polarized Fermi arc like surface states is observed at the interface between air and the Dirac metamaterials. Furthermore, eigen-reflection fields show the decoupling process from a Dirac point to two Weyl points. We also find the topological correlation between a Dirac point and vortex/vector beams in classical photonics. The

---

[*] xiangyuanjiang@126.com
[†] s.zhang@bham.ac.uk

**experimental feasibility of our scheme is demonstrated by designing a realistic metamaterial structure. The theoretical proposal of photonic Dirac point lays foundation for unveiling the connection between intrinsic physics and global topology in electromagnetism.**

Topological phases[1-4], first arising from Quantum Hall effect [5], have fundamentally revolutionized our understanding of the states of matter. In recent years, the novel concepts of topological phases have been transferred to photonic systems, leading to the discoveries of photonic analogy of Quantum Hall effect [6, 7], 2D/3D photonic topological insulators [8-12], Weyl and Dirac degeneracies [13-23]. These have remarkably enriched classical photonics, opening a path towards studying fundamental new states of light and possible revolutionary applications in edge/surface optics.

Dirac point is a four-fold band crossing defined in three dimensional momentum space, away from which energy band exhibits linear dispersion along arbitrary direction. As a central gapless topological phase, Dirac semimetal bridges conventional insulator, topological insulators and Weyl semimetals. By breaking either time reversal ($T$) or inversion symmetry ($P$), it may split into two Weyl points of opposite chirality. In photonics, four-fold degeneracy of Dirac point is usually achieved through various space symmetries in photonic crystals, where the Bloch modes, i.e. spatial degrees of freedom, play vital roles [11, 12, 21].

In this letter, by utilizing the intrinsic polarization degrees of freedom, we theoretically demonstrate a way to design photonic Dirac points in a medium with homogeneous effective electromagnetic properties, merely protected by electromagnetic duality symmetry, which is usually enabled by a fixed ratio between all the permittivity and permeability tensor elements [9, 12]. At the interface between air and the Dirac metamaterials, there exist spin-dependent topological surface states, which are the electromagnetic counterparts of Fermi arcs in electronic system. Because of the linearity of their k-space dispersion, the surface waves propagate nearly without diffraction, which is significant for the information transport and imaging applications. Because of the underlying topological relations, the topological charges discovered in classical photonics, such as vector/vortex beams, can be generated through the interaction of light with photonic Dirac points.

Here we first consider a uniaxial metamaterial with realistic resonant features in both permittivity and permeability. Specifically, the effective parameters of Dirac metamaterials are taking the form of $\varepsilon = diag\{\alpha,\alpha,\varepsilon_z\}, \mu = diag\{\varsigma,\varsigma,\mu_z\}$. For the sake of simplicity we assume $\alpha = \varsigma = constant$ and,

$$\varepsilon_z = 1 + f_1\omega_0^2/(\omega_0^2 - \omega^2), \mu_z = 1 + f_2\omega^2/(\omega_0^2 - \omega^2). \tag{1}$$

where $\omega_0$ indicates the resonance frequency and coefficients $f_1$ and $f_2$ are constants determined by the structure parameters which are adjustable. The presence of resonance along $z$ direction for both permittivity and permeability indicates that there exist two bulk plasmon modes, a longitudinal electric mode, and a longitudinal magnetic mode. From Eq. (1) it is straightforward to show that the degeneracy between the two longitudinal modes can be reached by setting $f_2 = 1 - 1/(1 + f_1)$.

The band structures for a specific set of parameters satisfying the above condition are shown in Fig. 1. Fig. 1(a) shows the plot of band structures in the $k_y$ - $k_z$ plane, where two bands are nearly overlapping with each other. There are two four-fold degeneracy points symmetrically displaced on the $k_z$ axis, as marked by the blue spheres. These two points are Dirac points that form the central focus of our study. Fig. 1(b) gives the dispersion relation of the material along $k_y$ and $k_z$ directions. In the $k_z$ direction, the two longitudinal modes and the two transverse modes are perfectly degenerate, respectively, across the whole frequency range. The band crossing between the two transverse modes and the two longitudinal modes is guaranteed by the orthogonality between them. Very recently, the crossing between a single longitudinal plasmon mode and one circular polarization mode in hyperbolic chiral metamaterials [23] and magnetized plasma [17] is found to show topological nature, exhibiting exotic properties of Weyl quasi-particles.

Fig. 1(c) shows the linear dispersions along $k_x$ or $k_y$ direction across the Dirac point, where bands are also degenerate in a frequency range around the Dirac point. The linear dispersions along all directions reveal a massless Dirac collective excitation. In the momentum space, the equi-frequency contour (EFC) at a frequency slightly below the Dirac points is shown in Fig. 1(d). One could see the EFC possesses double hyperbolas ($\varepsilon_z < 0, \mu_z < 0$), which is characteristic of a double hyperbolic metamaterial (DHM).

In the proposed Dirac metamaterial, the 3D Dirac points are protected by electromagnetic duality $\varepsilon = \eta\mu \ (\eta > 0)$ — an internal symmetry of the electromagnetic fields [24]. In the metamaterials described above $\eta = 1$. Under the following set of basis, $LCP(E = (\hat{x} - i\hat{y})/\sqrt{2}, H = iE)$, $L_1(E = \hat{z}, H = iE)$ and $RCP(E = (\hat{x} + i\hat{y})/\sqrt{2}, H = -iE)$, $L_2(E = \hat{z}, H = -iE)$, the Dirac Hamiltonian of the metamaterial takes a block-diagonal form consisting of two Weyl degeneracies [25],

$$H_D(k) = \begin{bmatrix} H_W(k) & \\ & H_W^*(k) \end{bmatrix}. \quad (2)$$

where the induced Weyl Hamiltonian, arising from the linear crossing between a longitudinal mode $L_1$ and a circularly polarized transverse mode RCP [17], is given by

$$H_W(k) = \begin{bmatrix} d_z k_z & d_x k_x - i d_y k_y \\ d_x k_x + i d_y k_y & 0 \end{bmatrix}, \quad (3)$$

where $d_i$ is the corresponding velocity determined by specific material parameters. Each Dirac point consists of two decoupled Weyl points with chirality of ±1 respectively as schematically shown in Supplemental material. [25]. The chirality of each Weyl node is determined by the intrinsic spin $s = \pm 1$ of the transverse mode involved, because both longitudinal modes are spinless $(s = 0)$.

Angle-resolved reflectance spectrum provides an intuitive way to show how a Dirac node can be decomposed into two decoupled complex-conjugation related Weyl points. As shown in Fig. 2(a), a plane wave with a specific polarization state is illuminated around the Dirac node (the blue points at the interface). Here we assume an ideal case that the overall system is under electromagnetic duality; the Dirac metamaterials $\varepsilon = \mu = \{k_D, k_D, 0\}$ with Dirac points is located at $x < 0$, while the $x > 0$ semi-space is occupied by an isotropic medium with $\varepsilon = \mu > k_D I$. Fig. 2(b) and (c) show the reflected polarization states and phases with RCP/LCP incidence, respectively. Obviously, circular polarizations are the eigenstates of the system due to the electromagnetic duality. In the momentum space, one anticlockwise loops around the Dirac point with RCP/LCP incidence acquires a $2\pi/-2\pi$ reflection phase leading to generation of vortex beam in reflection [25], which directly maps out the chirality of each constituent Weyl point of the Dirac degeneracy. With linear polarization incidence, the reflection field is symmetric about the $\Delta k_z$ axis as shown in Fig. 2 (d) and (e). This can be understood as the topological charge of a Dirac node being 0. Interestingly, radial and angular vector beams are generated by a Gaussian beam incidence with TE and TM polarization states, respectively (Fig. 2 (d) and (e)). Therefore, the above study shows that 3D photonic Dirac points are topologically linked to the vector/vortex photonics. Dirac point degeneracy provides a novel method in generating vector and vortex beams.

As mentioned above, Dirac point consists of two opposite Weyl points. A landmark of Weyl topological phase is the existence of Fermi arcs connecting projections of pairs of Weyl points. Therefore, a topologically nontrivial Dirac point may exhibit double Fermi arcs, which have been observed in electronic systems [26]. Here we calculate the topological surface states between the Dirac metamaterials and air, where air naturally has electromagnetic duality symmetry $(\varepsilon_{air} = \mu_{air} = I)$. Furthermore, owing to the transverse nature of electromagnetic waves, propagation of light in free space possesses Berry curvature in the momentum space given by $\Omega = \sigma k/|k|^3$, where $\sigma = \pm 1$ represents spin of RCP/LCP [27]. At the origin of momentum space, there exist singularities of Berry curvature, so called "Dirac monopoles". For LCP, the origin behaves like a sink of Berry flux, while for RCP it is the source. This nontriviality leads to the well-known spin-dependent Chern numbers 2σ. Thus, air provides topological nontriviality for the "Fermi arcs" connection [25].

As the duality symmetry between $\varepsilon$ and $\mu$ is preserved in both medium, at the interface formed by these two media, there exist two pairs of spin polarized surface states, as shown in Fig. 3(a). Each pair of surface states consists of spin up ($\sigma = +1$, indicated with pink surface) and spin down ($\sigma = -1$, indicated with cyan surface) surface states [25]. Fig. 3(b) shows EFC of the Fermi arc like surface states at Dirac points. From each Dirac point, there are two "Fermi arcs" connecting to the air light circle. In Fig. 3(c), EFCs at a frequency below Dirac points are presented. Interestingly, by setting $\varepsilon_z = \mu_z = \beta$ with $(\alpha > 1, \beta \leq 0)$, these "Fermi arcs" geometrically appear as straight line segments tangentially connecting air-circle and DHM-hyperbolas with linear dispersion relation in *k* space,

$$k_z = \pm \frac{\sqrt{\alpha^2 - 1}}{\sqrt{1 - \alpha\beta}} k_y \pm \frac{\sqrt{\alpha^2 - \alpha\beta}}{\sqrt{1 - \alpha\beta}} \qquad (4)$$

which leads to the near diffraction-less propagation of the surface state. The length of the surface dispersion segments is found to be strictly related to the decay constant in both air and DHM [25].

In order to demonstrate the topological protection and spin-momentum locking of the surface states, a full wave 3D simulation is performed using CST microwave studio with parameters of $\alpha = 3, \beta = 0$ (at the Dirac point frequency). The configuration is built by a cuboid-shaped Dirac metamaterials surrounded by air. A small port is positioned at the interface as the radiation source. Fig. 4 (a) and (b) show the backscattering immune transportation of spin-dependent surface waves helically bending around *z*-invariant sharp corners, indicating the one way surface state is right/left handed polarization [25]. These demonstrations confirm the robustness of spin "Fermi arcs" in the photonic Dirac metamaterials.

Having outlined the realization of Dirac point in metamaterials based on their effective parameters, we now demonstrate the experimental feasibility of our scheme by designing a realistic metamaterial structure. To simultaneously realize the required permittivity and permeability dispersion along *z* direction, we apply metallic helixes as shown in Fig. 5(a), where two mirrored helixes along *z* direction (clockwise and anticlockwise rotated helix-pair) to introduce both electric and magnetic resonances along z direction, while eliminating the chiral response. In the *x-y* plane, we apply $C_4$ rotation symmetry to achieve in-plane isotropic electromagnetic response. Thus, each unit cell consists of eight helixes in total. By tuning the geometric parameters of the structure, such as radius *R*, height *h*, period *P* along x/y direction and $p_z$ along z direction, the dispersion of $\varepsilon_z$ and $\mu_z$ can be adjusted leading to the degeneracy of the two longitudinal modes. In the microwave regime, the designed Dirac metamaterial can be readily fabricated [23, 28]. Fig. 5(b) and (c) present the dispersion curves of the bulk states of the proposed metamaterial along z and y directions, respectively. The presence of a four-fold crossing between two degenerate longitudinal modes (Fig. S5, L1 and L2) and two degenerate transverse modes (Fig. S5, T1 and T2) is observed [25]. Furthermore, the presence of Fermi arc like surface states is numerically confirmed as shown in Fig. 5(d), where a supercell consisting of 20 unit-cells surrounded by air is used in the surface state calculation. The localized surface states are shown in Fig. S6 [25].

The Dirac point here lies right at the critical transition between type I and type II. This is due to the flat dispersion relation of the bulk plasmon modes. For a type I Dirac point, the equifrequency surface (EFS) reduces to a single point in the momentum space, while for a type II Dirac point, the EFS consists of two touching cones. For the transitional Dirac point described here, the EFS is a straight line which corresponds to the degenerate electric and magnetic bulk plasmon modes. On the other hand, it is possible to introduce nonlocality into the system such that the bulk plasmon dispersion are not flat, but exhibiting either positive or negative dispersion [20]. That would lead to the construction of either type-I or type-II Dirac points. Note that type-II Dirac points in a photonic crystal were proposed in Ref. 21, where spatial degrees of freedom play a key role in constructing the state space around Dirac points. In contrast, the Dirac points discovered here exploits the internal degrees of freedom, i.e. the polarization states of electromagnetic wave, leading to very interesting polarization structures for light reflected around the Dirac points, as shown in Fig. 2.

Exploration of topological photonics protected by electromagnetic duality symmetry has introduced a number of intriguing optical phenomena [9, 12, 29]. However, the experimental observation still remains challenging, particularly in 3D up to now. It is expected that well developed metamaterials theory and finely designed high-permittivity meta-crystals [12] can provide an electromagnetic duality platform to discovery more new photonic topological phases.

In conclusion, by considering electromagnetic duality symmetry, we theoretically proposed the realization of photonic Dirac degeneracy and spin-dependent Fermi arc like surface states. This theoretical model is of great importance in understanding fundamental topological photonics, such as exhibiting transition from $Z_2$ topological insulator to Weyl degeneracies [4]. Furthermore, the fundamental link between photonic Dirac points and vector/vortex optics has been established through optical reflection with in-plane momentum matching the Dirac points.

**Acknowledgments** The work is partially supported by ERC Consolidator Grant (TOPOLOGICAL), the Royal Society and the Wolfson Foundation, the Leverhulme Trust (RPG-2012-674), and Horizon 2020 Action, Project No. 734578 (D-SPA). Q. G acknowledges the financial support by National Science Foundation of China (Grant No.11604216) and China Postdoctoral Science Foundation (Grant No.2016M600667). B. Y. acknowledges China Scholarship Council (201306110041). J. C. acknowledges support from the National Natural Science Foundation of China (Grant No.11574162). Y. X. acknowledges support from the National Natural Science Foundation of China (Grant No.61490713, 61505111).

Q. G. and B. Y. contributed equally to this work.

**References:**

[1] X.-L. Qi, and S.-C. Zhang, Rev. Mod. Phys. **83**, 1057 (2011).
[2] M. Z. Hasan, and C. L. Kane, Rev. Mod. Phys. **82**, 3045 (2010).
[3] C.-K. Chiu, J. C. Y. Teo, A. P. Schnyder, and S. Ryu, Reviews of Modern Physics **88**, 035005 (2016).
[4] N. P. Armitage, E. J. Mele, and A. Vishwanath, arXiv:1705.01111 (2017).
[5] K. v. Klitzing, G. Dorda, and M. Pepper, Physical Review Letters **45**, 494 (1980).
[6] F. D. M. Haldane, and S. Raghu, Physical Review Letters **100**, 013904 (2008).
[7] Z. Wang, Y. Chong, J. D. Joannopoulos, and M. Soljacic, Nature **461**, 772 (2009).
[8] M. C. Rechtsman, J. M. Zeuner, Y. Plotnik, Y. Lumer, D. Podolsky, F. Dreisow, S. Nolte, M. Segev, and A. Szameit, Nature **496**, 196 (2013).
[9] A. B. Khanikaev, S. Hossein Mousavi, W.-K. Tse, M. Kargarian, A. H. MacDonald, and G. Shvets, Nat. Mater. **12**, 233 (2013).
[10] W.-J. Chen, S.-J. Jiang, X.-D. Chen, B. Zhu, L. Zhou, J.-W. Dong, and C. T. Chan, Nature Communications **5**, 5782 (2014).
[11] L. Lu, C. Fang, L. Fu, S. G. Johnson, J. D. Joannopoulos, and M. Soljacic, Nat Phys **12**, 337 (2016).
[12] A. Slobozhanyuk, S. H. Mousavi, X. Ni, D. Smirnova, Y. S. Kivshar, and A. B. Khanikaev, Nat Photon **11**, 130 (2017).
[13] L. Lu, L. Fu, J. D. Joannopoulos, and M. Soljacic, Nat Photon **7**, 294 (2013).
[14] L. Lu, Z. Wang, D. Ye, L. Ran, L. Fu, J. D. Joannopoulos, and M. Soljačić, Science **349**, 622 (2015).
[15] W. Gao, M. Lawrence, B. Yang, F. Liu, F. Fang, B. Béri, J. Li, and S. Zhang, Physical Review Letters **114**, 037402 (2015).
[16] E. E. Narimanov, Faraday Discussions **178**, 45 (2015).
[17] W. Gao, B. Yang, M. Lawrence, F. Fang, B. Béri, and S. Zhang, Nat. Commun. **7**, 12435 (2016).
[18] H. Wang, L. Xu, H. Chen, and J.-H. Jiang, Physical Review B **93**, 235155 (2016).
[19] W.-J. Chen, M. Xiao, and C. T. Chan, Nat. Commun. **7**, 13038 (2016).
[20] M. Xiao, Q. Lin, and S. Fan, Phys. Rev. Lett. **117**, 057401 (2016).
[21] H.-X. Wang, Y. Chen, Z. H. Hang, H.-Y. Kee, and J.-H. Jiang, arXiv:1703.09899 (2017).
[22] J. Noh, S. Huang, D. Leykam, Y. D. Chong, K. P. Chen, and M. C. Rechtsman, Nat Phys **13**, 611 (2017).
[23] B. Yang, Q. Guo, B. Tremain, L. E. Barr, W. Gao, H. Liu, B. Béri, Y. Xiang, D. Fan, A. P. Hibbins, and S. Zhang, Nature Communications **8**, 97 (2017).
[24] I. Fernandez-Corbaton, X. Zambrana-Puyalto, N. Tischler, X. Vidal, M. L. Juan, and G. Molina-Terriza, Physical Review Letters **111**, 060401 (2013).
[25] See Supplemental Material online.
[26] S.-Y. Xu, C. Liu, S. K. Kushwaha, R. Sankar, J. W. Krizan, I. Belopolski, M. Neupane, G. Bian, N. Alidoust, T.-R. Chang, H.-T. Jeng, C.-Y. Huang, W.-F. Tsai, H. Lin, P. P. Shibayev, F.-C. Chou, R. J. Cava, and M. Z. Hasan, Science **347**, 294 (2015).
[27] K. Y. Bliokh, D. Smirnova, and F. Nori, Science **348**, 1448 (2015).
[28] W. Bingnan, Z. Jiangfeng, K. Thomas, K. Maria, and M. S. Costas, Journal of Optics A: Pure and Applied Optics **11**, 114003 (2009).
[29] J.-W. Dong, X.-D. Chen, H. Zhu, Y. Wang, and X. Zhang, Nature materials **16**, 298 (2017).

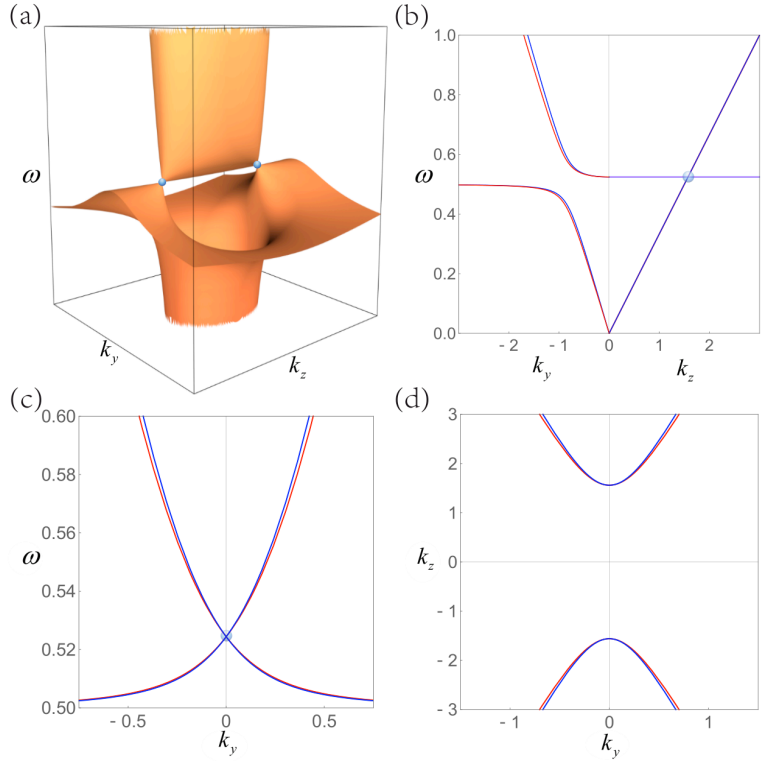

Fig. 1. Band structure of bulk states with $\alpha = \varsigma = 3$. (a) Effective bulk band structure on the $k_y$-$k_z$ plane with the Dirac points marked by blue sphere. (b) The dispersion relation along $k_y$ and $k_z$ direction respectively, while (c) gives the dispersion along $k_y$ direction with $k_z$ fixed at the Dirac point. In both (b) and (c) the Dirac Points are marked with the blue points. (d) Equi-frequency contour at a frequency below Dirac points, which indicates the hyperbolic property of the material corresponds to the Dirac cone.

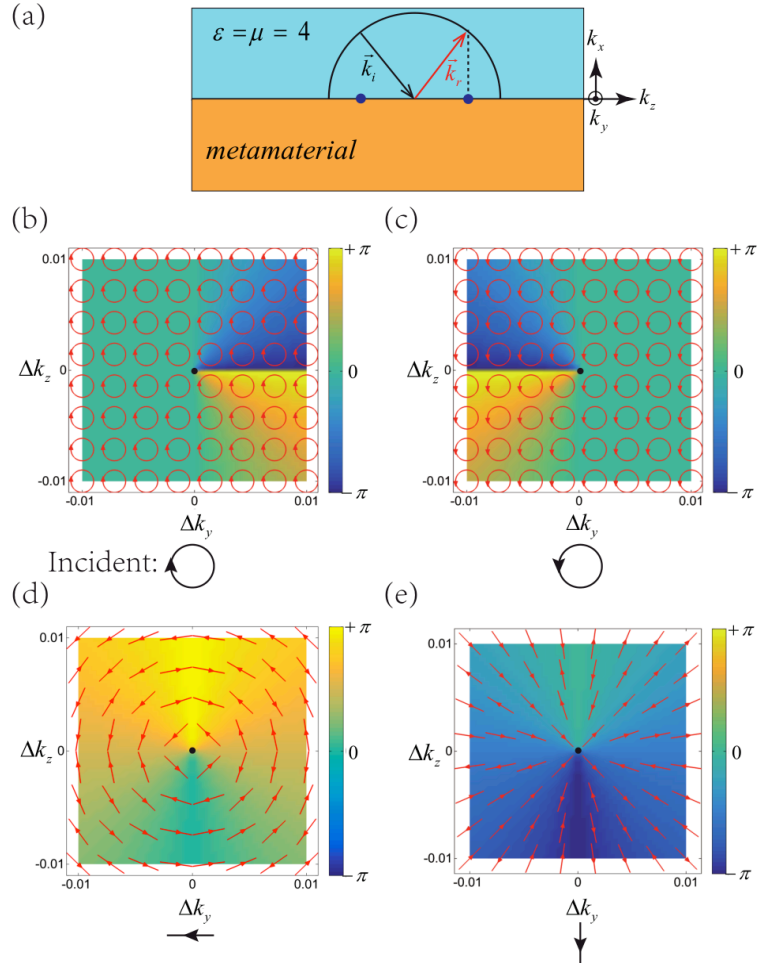

Fig. 2. Reflection spectrum around a Dirac point. (a) Configuration of the reflection calculatioin and Dirac points (two blue dots) in momentum space. One plane wave is incident $(k_i)$ from the backgroud material $x > 0$ and totally reflected $(k_r)$ back at the interface as indicated. (b) and (c) show the reflected E-field with respect of right/left circular polarization incidence, respectively. (d) and (e) are similar to (b) and (c) with the incidences of two orthogonal linear polarizations. The incident polarization state is illustrated on the bottom of each panel.

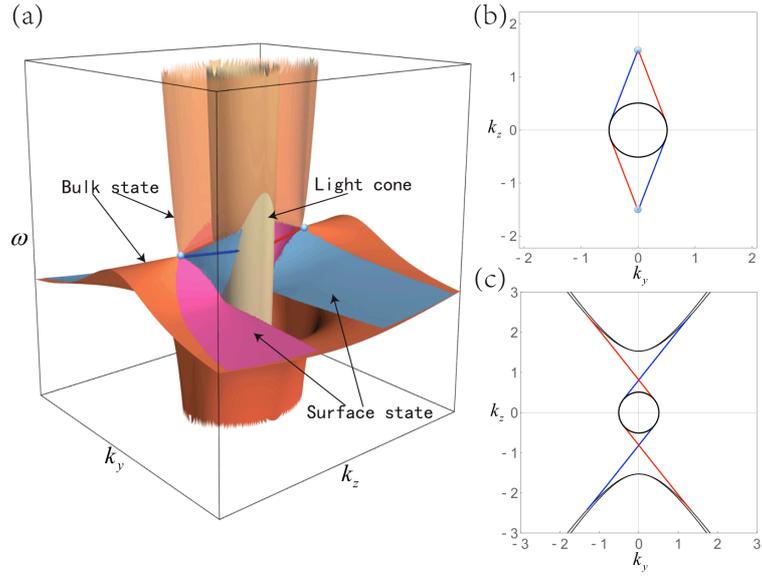

Fig. 3. Bulk and surface states of the effective medium, (a) 3D band structures of bulk and surface states of the effective medium. Spin-up (Spin-down) surface state between two Dirac points and vacuum is indicated by the red (blue) surface. The blue and red lines highlight the photonic "Fermi arcs" at Dirac point. Equi-frequency contours at two different frequencies could be seen in (b) and (c) corresponding to the frequency at Dirac point and below it, respectively. The red and blue lines in (b) and (c) represent the spin up and spin down topological surface states, while the black lines represent the effective material and vacuum bulk states.

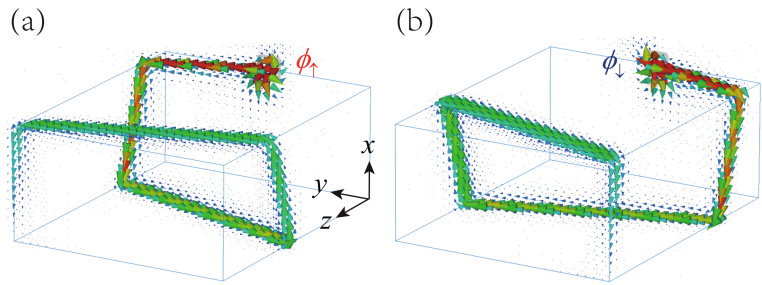

Fig. 4. Surface states indicated by power flow simulated in 3D by CST time domain. The (a) spin-up and (b) spin-down surface states propagate helically along z direction clockwise and anti-clockwise respectively, on the surface of cuboid effective material which is capsulated by vacuum.

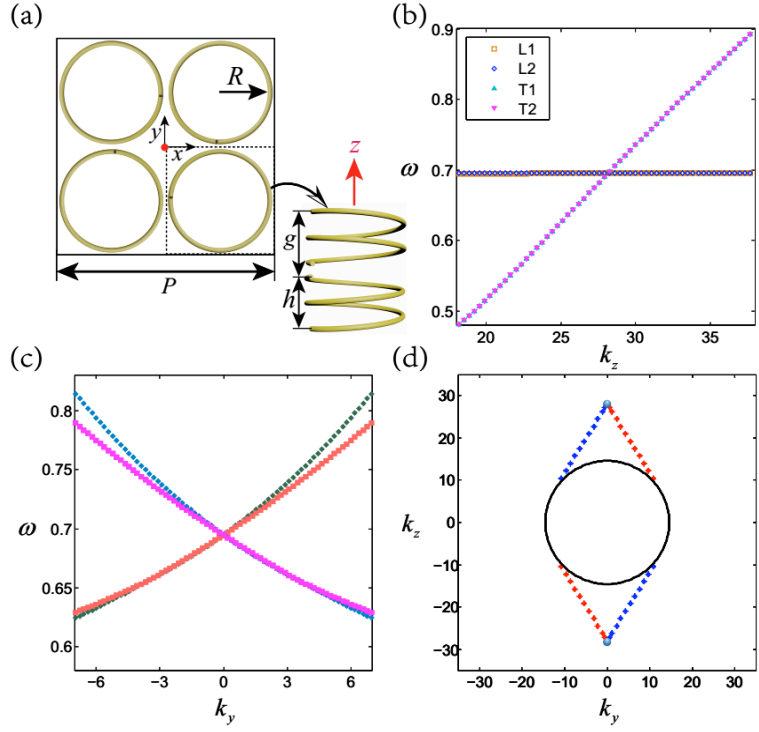

Fig. 5 Dirac points with realistic metamaterial structure. (a) The metamaterial structure with top view of one unit cell composed with $C_4$ symmetry distributed helix-pairs, and the front view of a pair of mirrored helixes with opposite rotation direction with parameters $R$=20mm, $p_x$=$p_y$=$P$=90mm, $p_z$=50mm, $g$=25mm, $h$=20mm. (b) and (c) are the simulated band strucature of the metamaterial along $k_z$ and $k_y$ direction respectively. (d). Surface state dispersion at Dirac poin frequency (0.695GHz).